\documentclass[12pt]{article}
\usepackage[dvips]{graphicx}
\usepackage[american]{babel}
\usepackage{amsmath}

\newtheorem{theorem}{\bfseries Theorem}

\begin{document}
\title{Randomly evolving trees III}
\author{L. P\'al, \\\footnotesize{KFKI Atomic Energy Research Institute
1525 Budapest, P.O.B. 49. Hungary}}

\date{June 20, 2003}

\maketitle

\begin{abstract}
{\footnotesize The properties of randomly evolving special trees
having defined and analyzed already in two earlier papers
(arXiv:cond-mat/0205650 and arXiv: cond-mat/0211092) have been
investigated in the case when the continuous time parameter
converges to infinity. Equations for generating functions of the
number of nodes and end-nodes in a stationary (i.e. infinitely
old) tree have been derived. In order to solve exactly these
equations we have chosen three different distributions for the
number of new nodes $\nu$ produced by one dying node. By using
appropriate method we have calculated step-by-step the
probabilities of finding $n=1, 2, \ldots $ nodes as well as
end-nodes in a stationary random tree. Analyzing the results of
numerical calculations we have observed that the qualitative
properties of stationary random trees depend hardly on the
character of distribution of $\nu$. The conclusion to be correct
that in the evolution process the formation of a rod-like
stationary tree is much more probable than the formation of a tree
with many branches. We have established that the probability of
finding $n$ nodes in a stationary tree depends sensitively on the
average value of $\nu$ and has a maximum the location of which is
increasing with $n$ but remains always smaller than unity. This is
also true for the end-nodes.}

\vspace{0.2cm}


\end{abstract}

\section{Introduction}

In previous two papers \cite{lpal0205}, \cite{lpal0211} we defined
and analyzed random processes with continuous time parameter
describing the evolution of special trees consisting of {\em
living and dead nodes} connected by {\em lines}. It seems to be
appropriate to repeat briefly the characteristic features of the
evolution process. The initial state ${\mathcal S}_{0}$ of the
tree corresponds to a single living node called {\em root} which
at the end of its life is capable of producing $\nu = 0, 1, \dots$
new living nodes, and after that it becomes immediately dead. If
$\nu > 0$ then the new nodes are promptly connected to the dead
node and each of them {\em independently of the others} can evolve
further like a root. The random evolution of trees with continuous
time parameter has not been investigated intensively recently. The
main interest since the late 1990s has been focussed on the study
of non-equilibrium networks \cite{dorog03} occurring in common
real world. The evolution mechanism of trees with living and dead
nodes may be useful in some of biological systems where the
branching processes are dominant.

In what follows, we will use notations applied in \cite{lpal0205}
and \cite{lpal0211}. It seems to be useful to cite the basic
definitions. The probability to find the number $\nu$ of living
nodes produced by one dying precursor equal to $j$ was denoted by
$f_{j}$ where $j \in {\mathcal Z}$.~\footnote{${\mathcal Z}$ is
the set of non-negative integers.} For the generating function as
well as the expectation value and the variance of $\nu$ we used
the following notations:
\[{\bf E}\{z^{\nu}\} = q(z), \;\;\;\;\;\;
{\bf E}\{\nu\} = q_{1} \;\;\;\;\;\; \mbox{and} \;\;\;\;\;\; {\bf
D}^{2}\{\nu\} = q_{2} + q_{1} - q_{1}^{2}, \]  where
\[ q_{j} =  \left[\frac{d^{j} q(z)}{dz^{j}}\right]_{z=1},
\;\;\;\;\;\; j = 1, 2, \ldots \] are {\em factorial moments} of
$\nu$. It was shown in \cite{lpal0205} that the time dependence of
the random evolution is determined almost completely by the
expectation value $q_{1}$. In accordance to this the evolution was
called subcritical if $q_{1}<1$, critical if $q_{1} = 1$ and
supercritical if $q_{1}>1$.

In the further considerations we are going to use four
distributions for the random variable $\nu$. As shown in
\cite{lpal0205} the equations derived for the first and the second
moments of the number of nodes are {\em independent of the
detailed structure of the distribution of $\nu$} provided that the
moments $q_{1}$ and $q_{2}$ are finite. We called distributions of
this type for $\nu$ {\em arbitrary} and used the symbol $\nu \in
{\bf a}$ for its notation . In many cases it seems to be enough to
apply the truncated distribution of $\nu$. If the possible values
of the random variable $\nu$ are $0, 1$ and $2$ with probabilities
$f_{0}, f_{1}$ and $f_{2}$, respectively, then in the previous
paper \cite{lpal0211} the distributions of this type were denoted
by $\nu \in {\bf t}$. Many times it is expedient to assume
distributions to be completely determined by one parameter. As
known the geometric and Poisson distributions are such
distributions. In paper \cite{lpal0211} we used the symbols $\nu
\in {\bf g}$ and $\nu \in {\bf p}$ to identify these
distributions.

The distribution function of the lifetime $\tau$ of a living node
will be supposed to be exponential, i.e. ${\mathcal P}\{\tau \leq
t\} = 1 - e^{-Qt}$. In order to characterize the tree evolution
two non-negative integer valued random functions $\mu_{\ell}(t)$
and $\mu_{d}(t)$ are introduced: $\mu_{\ell}(t)$ is the number of
living nodes, while $\mu_{d}(t)$ is that of dead nodes at $t \geq
0$. The total number of nodes at $t \geq 0$ is denoted by
$\mu(t)$.

Clearly, the nodes can be sorted into groups according to {\em the
number of outgoing lines}. Following the notation in
\cite{lpal0211} the number of nodes with $k \geq 0$ outgoing lines
at time instant $t \geq 0$ is denoted by $\mu(t, k)$. A node not
having outgoing line is called {\em end-node}. It is obvious that
an end-node is either live or dead. Therefore, the number of
end-nodes $\mu(t, 0)$ can be written as a sum of numbers of living
and dead end-nodes, i.e. $ \mu(t, 0) = \mu_{\ell}(t, 0) +
\mu_{d}(t, 0) = \mu_{0}(t)$. Since all living nodes are end-nodes
$\mu_{\ell}(t, 0)$ can be replaced by $\mu_{\ell}(t)$. The total
number of dead nodes $\mu_{d}(t)$ is given by $\mu_{d}(t) =
\sum_{k=0}^{\infty} \mu_{d}(t, k)$.

In this paper we are dealing with properties of $\mu(t), \;
\mu_{\ell}(t)$ and $\mu_{0}(t)$ when $t \rightarrow \infty$. We
will call the random trees arising from a single root after
elapsing infinite time {\em stationary}.

In Section $2$ the basic properties of probability distributions
of the number of nodes, living and end-nodes are investigated when
$t \rightarrow \infty$. Special attention is paid in Section $3$
to the effect of distribution law of the number of outgoing lines.
Three different distributions of $\nu$ are investigated. In order
to simplify the notation, indices referring to different
distributions of $\nu$ are usually omitted in formulas. Finally,
the characteristic properties of stationary random trees are
summarized in Section $4$.

\section{General considerations}

Let us introduce the notion of {\em tree size} which is nothing
else but the total number of nodes $\mu(t)$ at time moment $t \geq
0$. We want to analyze the asymptotic behavior of the tree size,
i.e. the behavior of the random function $\mu(t)$ when $t
\rightarrow \infty$. We say the {\em limit random variable}
\[ \tilde{\mu} \stackrel{\mathrm{d}}{=} \lim_{t \rightarrow \infty} \mu(t), \]
exists in the sense that the relation:
\begin{equation} \label{1}
\lim_{t \rightarrow \infty} {\mathcal P}\{\mu(t) = n|{\mathcal
S}_{0}\} = {\mathcal P}\{\tilde{\mu} = n\}
\end{equation}
is true for all positive integers $n$, where ${\mathcal S}_{0}$
denotes the initial state of the tree. A randomly evolving tree is
called {\em "very old"} when $t \Rightarrow \infty$, and a very
old tree, as mentioned already, will be named {\em stationary
random tree}.

It is elementary to prove that if the limit probability ${\mathcal
P}\{\tilde{\mu} = n\} = p_{n}$ exists, then the generating
function
\begin{equation}
\label{2} g(z) = \sum_{n=1}^{\infty} p_{n}\;z^{n}
\end{equation}
is determined by one of the fixed points of the equation
\begin{equation}
\label{3}  g(z) = z\;q[g(z)].
\end{equation}
It can be shown that if $q_{1} \leq 1$, then the fixed point to be
chosen has to satisfy the limit relation
\begin{equation}
\label{4} \lim_{z \uparrow 1} g(z) = 1,
\end{equation}
while if $q_{1} > 1$, then it should have the property
\begin{equation}
\label{5} \lim_{z \uparrow 1}g(z) < 1
\end{equation}
and independently of $q_{1}$ the equation $g(0) = 0$ must hold.
The relation (\ref{4}) means that the probability to find
stationary tree of finite size is evidently $1$, if $q_{1} \leq
1$, but if $q_{1} > 1$, then
\[ \lim_{z \uparrow 1} g(z) = \sum_{n=1}^{\infty}p_{n} < 1, \]
i.e. the probability to find a stationary tree of infinite size is
equal to
\[ w_{\infty} = 1 - \sum_{n=1}^{\infty}p_{n}. \]
The proof of the proposition is simple. Let us assume that
\[ q(z) = \sum_{n=0}^{\infty} f_{n} z^{n} \]
is a {\em probability generating function}, i.e.  $q(1) =
\sum_{n=0}^{\infty} f_{n} = 1$ and $f_{1} \neq 1$. We need the
following theorem:
\begin{theorem}
\label{th1} If $q'(1) = q_{1} \leq 1$, then $q(z) > z, \;\;
\forall\; 0 \leq z < 1$; while if $q'(1) = q_{1} > 1$, then there
is a point $0< z_{0} < 1$ such that $q(z_{0}) = z_{0}$, i.e. $q(z)
> z, \;\; \mbox{if} \;\; 0 < z < z_{0}\;$
and $\;q(z) < z, \;\; \mbox{if} \;\; z_{0} < z < 1$.
\end{theorem}

Let us introduce the function $\varphi(z) = q(z) - z$. Since
$q(z)$ is convex, i.e. all derivatives are positive in the
interval $[0, 1]$ it is evident that
\[ \frac{d\varphi(z)}{dz} = \varphi'(z)\]
is a nondecreasing function of $z,\;\; \forall\;\; z \in [0,1]$.
If $q'(1) < 1$, then $\varphi'(z) < 0$, i.e. $\varphi'(z)$ is
nondecreasing, negative valued function of $z$ in $0 \leq z < 1$.
Since $\varphi(1) = 0$ it is obvious that $\varphi(z) < 0$, if $0
< z < 1$, i.e. $q(z) < z$, and this is the first statement of
Theorem \ref{th1}. If $q'(1) > 1$, then $\varphi'(z) > 0$, and
since $\varphi(1) = 0$ the inequality $\varphi(z) < 0$ has to be
true for all $z < 1$ lying near $1$. On the other side $\varphi(0)
= f_{0} > 1$, hence there should exist one ~\footnote{Existence of
more than one $z_{0}$ is excluded because $q(z)$ is
convex.}$z_{0}$ in $(0,1)$ which satisfies the equation
$\varphi(z_{0}) = 0$ and that implies the second statement of
Theorem \ref{th1}.

It seems to be important to investigate {\em how the living nodes
behave} in very old, i.e. stationary trees. Intuitively one can
say that stationary trees arising in subcritical evolution do not
contain living nodes and they have finite average size. At the
same time, it seems to be quite obvious that stationary trees
originating in supercritical evolution could have living nodes
with non-zero probability, i.e. they are entities of {\em "eternal
life"}.

In order to give more precise answer the generating function of
the random variable
\[ \tilde{\mu}_{\ell} \stackrel{\mathrm{d}}{=} \lim_{t \rightarrow
\infty} \mu_{\ell}(t) \] should be derived. It is easy to show
that
\[  {\bf E}\{z ^{\tilde{\mu}_{\ell}}\} = g^{(\ell)}(z) =
\sum_{n=0}^{\infty} p_{n}^{(\ell)} z^{n} \] is one of the fixed
points of the equation
\begin{equation}
\label{6}  q\left[g^{(\ell)}(z)\right] = g^{(\ell)}(z).
\end{equation}
By Theorem $1$, equation (\ref{6}) has one fixed point if $q_{1} <
1$, namely $g_{1}^{(\ell)}(z) = 1, \;\;\; \forall \; z \in [0,1]$,
and from this it follows that
\[ p_{n}^{(\ell)} = \left\{\begin{array}{ll}
1, & \mbox{if $n=0$,} \\
\mbox{ } & \mbox{ } \\
0, & \mbox{if $n>0$,} \end{array} \right. \] i.e. the probability
that a subcritical very old tree does not have living node, is
exactly $1$. If $q_{1} > 1$, then besides $g_{1}^{(\ell)}(z) = 1$
equation (\ref{6}) has another fixed point given by
$\;g_{2}^{(\ell)}(z) < 1, \;\; \forall \; z \in [0,1]$, hence the
probability ${\mathcal P}\{\tilde{\mu}_{\ell} > 0 \} = 1 -
g_{2}^{(\ell)}(0)$ should be larger than zero. In other words, a
supercritical stationary tree may evolve infinitely with certain
non-zero probability.~\footnote{More precise formulation would be
the following: in countable set ${\mathcal T}$ of subcritical
stationary trees the measure of subset containing trees with
living nodes is zero, while in that of supercritical trees the
measure of subset consisting of trees with living nodes is larger
than zero.}

The expectation value and the standard deviation of the total
number of nodes can be used to characterize the size of a
stationary random tree. From (\ref{3}), one obtains
\[ {\bf E}\{\tilde{\mu}\} = \left[\frac{dg}{dz}\right]_{z
\uparrow 1} = \frac{1}{1-q_{1}}, \;\;\;\;\;\; \mbox{if} \;\;\;\;\;
q_{1} < 1. \] In a supercritical evolution, i.e. when  $q_{1} >
1$, the expectation value ${\bf E}\{\tilde{\mu}\}$ does not exist,
but the limit relation
\[ \lim_{t \rightarrow \infty}{\bf E}\{\mu(t)\;e^{-(q_{1}-1)Qt}\}
= \frac{q_{1}}{q_{1}-1} \] can be simply proved. We note here that
in the critical case, i.e. when $q_{1}=1$, the average tree size
becomes infinite linearly, i.e. we have the relation $\lim_{t
\rightarrow \infty}{\bf E}\{ \mu(t)/Qt\} = 1$.

The standard deviation of the total number of nodes in stationary
random trees is given by
\[ {\bf D}\{\tilde{\mu}\} = \frac{{\bf D}\{\nu\}}{(1-q_{1})^{3/2}},
\;\;\;\;\;\; \mbox{if} \;\;\;\;\; q_{1} < 1. \] In a supercritical
state the standard deviation does not exist, but it can be readily
shown that
\[ \lim_{t \rightarrow \infty} {\bf
D}\{\mu(t)\;e^{-2(q_{1}-1)Qt}\} = \frac
{q_{1}}{\sqrt{q_{1}-1}}\;\left(1 + \frac{{\bf
D}\{\nu\}}{q_{1}-1}\right), \;\;\;\; \mbox{if} \;\;\;\; q_{1} > 1.
\]

Finally, we would like to deal briefly with some of the {\em
properties of end-nodes} in stationary random trees. By using
equation ($14$) of \cite{lpal0211}, it can be shown that the
generating function
\[ g_{0}(z) = {\bf E}\{z^{\tilde{\mu}_{0}}\} = \sum_{n=0}^{\infty}
p_{n}^{(0)} z^{n}, \;\;\;\;\;\; \mbox{where} \;\;\;\;\;\;
\tilde{\mu}_{0} \stackrel{\mathrm{d}}{=} \lim_{t \rightarrow
\infty} \mu(t,0) \] is nothing else than one of fixed points of
the following simple equation:
\begin{equation}
\label{7}  q\left[g_{0}(z)\right] = g_{0}(z) + (1-z)\;f_{0}.
\end{equation}
Substituting $z = 0$ we obtain
\[ g_{0}(0)\;\left\{1 -
\sum_{n=1}^{\infty}f_{n}\;[g_{0}(0)]^{n}\right\} = 0, \] i.e.
$g_{0}(0) = 0$, hence ${\mathcal P}\{\tilde{\mu}_{0} = 0\} = 0$.

In the sequel we will investigate the properties of stationary
trees when $q(z)$ is known. In this case we can obtain exact
expressions for probabilities ${\mathcal
P}\{\tilde{\mu}=n\},\;\;{\mathcal P}\{\tilde{\mu}_{\ell}=n\},
\;\;{\mathcal P}\{\tilde{\mu}_{0}=n\}, \;\; \forall
n=1,2,\ldots\;$ from the corresponding generating function.

\section{Known distribution of $\nu$}

\subsection{Truncated arbitrary distribution of $\nu$}

The generating function of $\nu$ is given by
\begin{equation}
\label{8} q(z) = f_{0} + f_{1}\;z + f_{2}\;z^{2} = 1 +
q_{1}\;(z-1) + \frac{1}{2}q_{2}\;(z-1)^{2}
\end{equation}
with the restriction for $q_{1}$ and $q_{2}$ determined by the
equality $f_{0} + f_{1} + f_{2} = 1$. (See Fig. 1 in
\cite{lpal0211}.) By using Eq. (\ref{3}) and applying (\ref{8}) we
have
\[ z\;\left[1 + q_{1}\;(g - 1) +
\frac{1}{2}q_{2}\;(g - 1)^2\right] = g. \] The fixed point of this
equation is
\begin{equation}
\label{9} g(z) = 1 + \frac{1}{q_{2}z}\;\left[1 - q_{1} z -
\sqrt{(1 - q_{1} z)^{2} + 2q_{2}z(1 - z)}\right],
\end{equation}
is also the generating function of $\tilde{\mu}$. It is an
elementary task to show that
\begin{equation}
\label{10}  g(z) = \left\{ \begin{array}{ll}
0, & \mbox{if $z =0$,} \\
1, & \mbox{if $z =1$, and $q_{1} \leq 1$} \\
1 - 2(q_{1} - 1)/q_{2}, & \mbox{if $z =1$, and $q_{1} > 1$.}
\end{array} \right.
\end{equation}

\begin{figure} [ht!]
\protect \centering{\includegraphics[height=8cm,
width=12cm]{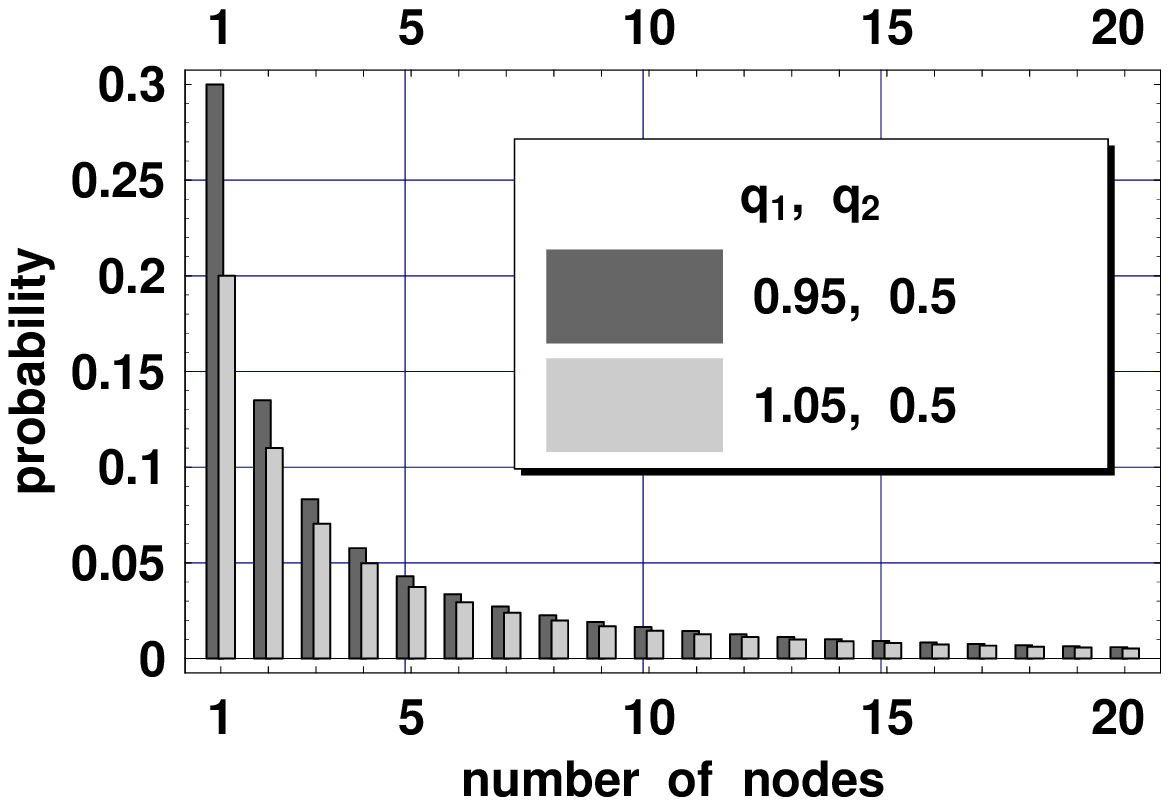}}\protect
\caption{\footnotesize{Probabilities to find $n=1, 2, \ldots$
nodes in a "very old" ($t=\infty$) tree when the distribution of
$\nu$ is truncated arbitrary. The dark and light bars correspond
to subcritical and slightly supercritical tree evolutions,
respectively.}} \label{fig1}
\end{figure}
Since
\[ g(z) = \sum_{n=1}^{\infty} p_{n}\;z^{n}, \]
performing expansion of generating function (\ref{9}) into power
series of $\;z$, we can determine the probabilities $\;\;p_{n},
\;\; \forall \; n \in {\mathcal Z}$ easily.  After elementary
calculations we obtain
\begin{equation} \label{11}
p_{n} =
\frac{1}{2\sqrt{\pi}q_{2}}\left[\frac{\Gamma(n+1/2)}{\Gamma(n+2)}\;\left(U^{n+1}
+ V^{n+1}\right) - W_{n+1}\right],
\end{equation}
where
\[ U = q_{1}-q_{2} + \sqrt{2q_{2}}\;\sqrt{1 - q_{1} + \frac{1}{2}q_{2}}, \]
\[ V = q_{1}-q_{2} - \sqrt{2q_{2}}\;\sqrt{1 - q_{1} + \frac{1}{2}q_{2}}, \]
and
\[ W_{n+1} =
\frac{1}{4\pi}\;\sum_{j=1}^{n}\frac{\Gamma(j-1/2)}{\Gamma(j+1)}
\;\frac{\Gamma(n-j+1/2)}{\Gamma(n-j+2)}\;U^{j}\;V^{n+1-j}.\]

\begin{figure} [ht]
\protect \centering{\includegraphics[height=8cm,
width=12cm]{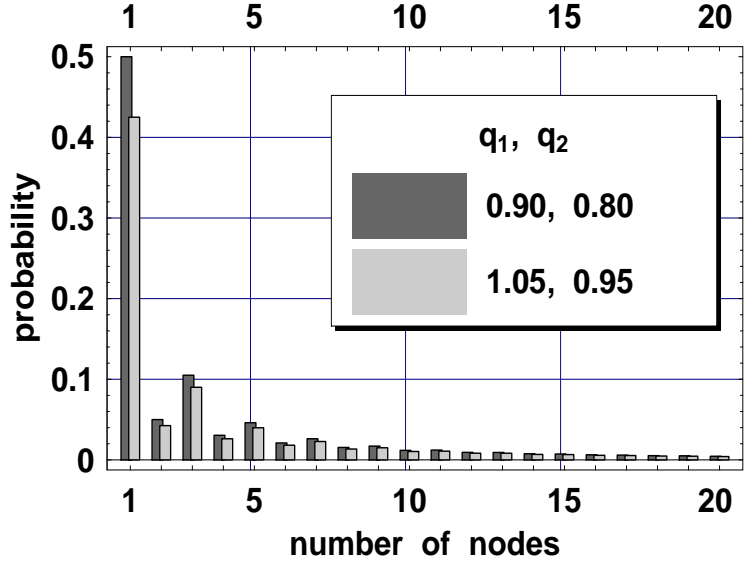}}\protect
\caption{\footnotesize{Probabilities to find $n=1, 2, \ldots$
nodes in a "very old" ($t=\infty$) tree when the distribution of
$\nu$ is truncated arbitrary and $f_{1}=0.1$. The dark and light
bars correspond to subcritical and slightly supercritical tree
evolutions, respectively.}} \label{fig2}
\end{figure}

The probabilities $p_{n}, \;\;\; n=1,2,\ldots$ can be seen in Fig.
\ref{fig1}. These are the probabilities to find $n=1,2,\ldots$
nodes in a "very old" tree developed according to the distribution
of $\nu \in {\bf a}$. The dark and light bars correspond to
subcritical $(q_{1}=0.95, q_{2}=0.5)$ and slightly supercritical
$(q_{1}=1.05, q_{2}=0.5)$ tree evolutions, respectively. In the
last case $w_{\infty} = 0.2$ and it is not surprising that the
probabilities $p_{n}$ for finite $n$ are larger in sub- than in
supercritical trees.

Fig. \ref{fig2} shows the dependence of $p_{n}$ on $n$ in sub- and
supercritical evolutions, respectively. If the difference $q_{1} -
q_{2} = f_{1}$ is small enough, let us say $0.1$, then one
observes a special phenomenon, namely, an {\em "oscillation"} in
the dependence of $p_{n}$ versus $n$, what is clearly seen in Fig.
\ref{fig2}.

\begin{figure} [ht!]
\protect \centering{\includegraphics[height=6cm,
width=9cm]{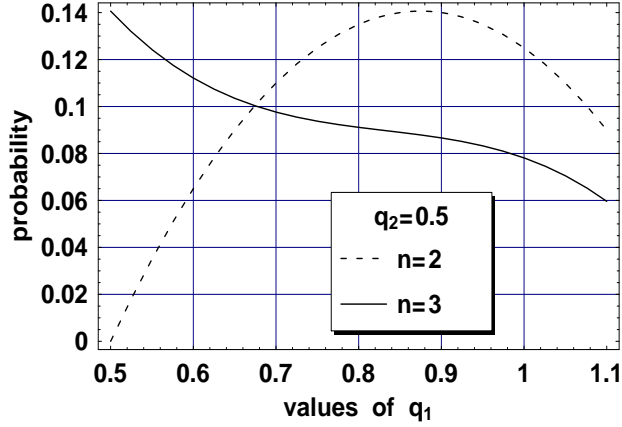}}\protect
\caption{\footnotesize{Dependence of probabilities $p_{n}, \;\;
n=2,3$ on $q_{1}$ at $q_{2}=0.5$}} \label{fig3}
\end{figure}

In order to explain the origin of "oscillation"  we calculated the
dependence of probabilities $p_{2}$ and $p_{3}$ on $q_{1}$ at
$q_{2}=0.5$. In Fig. \ref{fig3}, we can see that $p_{3}
> p_{2}$ in the interval $0 \leq q_{1} < q_{1}^{(c)}(2)$. For the
sake of simplicity the upper limit $q_{1}^{(c)}(2)$ will be called
"critical $q_{1}$".

The reason of oscillation is trivial. Clearly, if $q_{1} - q_{2} =
f_{1} = 0$, then there are no random trees having nodes of even
number, i.e. $p_{2j}=0, \;\; j=1,2, \ldots$, therefore, it should
be an interval $0 \leq f_{1} < f_{1}^{(c)}(2j)$ in which
\[ p_{2j} <  p_{2j+1} \;\;\;\;\;\;  \forall\; j=1, 2, \ldots. \]

We calculated the critical values of $q_{1}$ corresponding to
$f_{1}^{(c)}(2j)$ at $q_{2}=0.4(0.1)1.0$. The results are shown in
Table 1. As expected $q_{1}^{(c)}(2j)$ is slightly decreasing with
$j$ when $q_{2}$ is fixed.

\begin{table}[ht!]
\begin{center}
Table $1$: {\footnotesize Critical values of $q_{1}^{(c)}(2j)$ at
$q_{2} = 0.4(0.1)1.0$.}
\end{center}
\begin{center}
\begin{tabular}{|c|c|c|c|} \hline
$q_{2}$ & $p_{2}=p_{3}$& $p_{4}=p_{5}$ & $p_{6}=p_{7}$ \\
\hline 0.4 & 0.553 & 0.514 & 0.491 \\
\hline 0.5 & 0.674 & 0.634 & 0.609 \\
\hline 0.6 & 0.789 & 0.749 & 0.723 \\
\hline 0.7 & 0.897 & 0.858 & 0.833 \\
\hline 0.8 & 0.100 & 0.963 & 0.938 \\
\hline 0.9 & 1.098 & 1.062 & 1.039 \\
\hline 1.0 & 1.191 & 1.157 & 1.135 \\
\hline
\end{tabular}
\label{tab1}
\end{center}
\end{table}

At this point it is worthwhile to underline one of the most
characteristic properties of stationary random trees. As known,
all {\em moments} of the total number of nodes are converging to
infinity when the evolution is supercritical and $t \rightarrow
\infty$. However, the {\em probability} to find a "very old"
(stationary) supercritical tree of finite size is always larger
than zero and, therefore, the probability to find a supercritical
tree of infinite size is $w_{\infty} = 1 - \sum_{n=1}^{\infty}$,
and is always smaller than one.

In the case of truncated arbitrary distribution of $\nu$ one can
write Eq. (\ref{10}) in the form:
\begin{equation}
\label{12}  \sum_{n=1}^{\infty} p_{n} = \left\{ \begin{array}{ll}
      1, & \mbox{if $q_{1} \leq 1$,} \\
            1 - 2(q_{1}-1)/q_{2}, & \mbox{if $q_{1} > 1$,}
                    \end{array} \right.
\end{equation}
and so, one can formulate the statement as follows: a
supercritical "very old" tree may be finite with probability $1 -
2(q_{1}-1)/q_{2}$, and consequently, infinite with probability
$w_{\infty} = 2(q_{1}-1)/q_{2}$. It is elementary to prove that
$2(q_{1}-1)/q_{2} \leq 1$,  if $q_{1} > 1$.

Let us get some insight into the behavior of living nodes in
stationary random trees. From Eq. (\ref{6}) we obtain  for all
$|z| \geq 0$ that
\[ g_{1,2}^{(\ell)} = \left\{ \begin{array}{ll}
1, & \mbox{if $q_{1} < 1$,} \\
f_{0}/f_{2} = 1-2(1-q_{1})/q_{2}, & \mbox{if $q_{1} > 1$,}
\end{array} \right. \]
and from this we can conclude~\footnote{If $q_{1} < 1$, then
$f_{0} > f_{2}$ and hence the probability $p_{0}^{(\ell)}$ must
have the value $1$. If $q_{1} > 1$, then $f_{0} < f_{2}$ and it
means that the probability $p_{0}^{(\ell)}$ should be equal to
$f_{0}/f_{2}$.} that
\begin{equation}
\label{13} p_{0}^{(\ell)} = \left\{ \begin{array}{ll}
1, & \mbox{if $q_{1} < 1$,} \\
f_{0}/f_{2} = 1-2(1-q_{1})/q_{2}, & \mbox{if $q_{1} > 1$.}
\end{array} \right.
\end{equation}
It seems to be appropriate  to underline the notion of the second
half of Eq. (\ref{13}). It is expressing that the probability of
finding living nodes in supercritical stationary random tree is
$1-f_{0}/f_{2}$, i.e.
\[ {\mathcal P}\{ \tilde{\mu}_{\ell} > 0\} =
2\;\frac{q_{1}-1}{q_{2}}. \]

As to the {\em probabilistic properties of end-nodes} in
stationary trees we have to go back to Eqs. (\ref{7}) and
(\ref{8}). It is elementary to show that the generating function
of $\tilde{\mu}_{0}$ is nothing else than
\begin{equation}
\label{14} g_{0}(z) = 1 + \frac{1}{2f_{2}}\;\left[1 - f_{1} -
2f_{2} - (1 - f_{1})\;\sqrt{1 - \frac{4f_{2}\;f_{0}}{(1 -
f_{1})^{2}}\;z}\;\right].
\end{equation}
It is seen immediately that
\[ \lim_{z \uparrow 1} \tilde{g}_{0}(z) = \tilde{g}_{0}(1) =
\left\{ \begin{array}{ll} 1, & \mbox{if $q_{1} \leq 1$,} \\
1 - 2(q_{1}-1)/q_{2}, & \mbox{if $q_{1} > 1$,}
\end{array} \right. \]
\begin{figure} [ht!]
\protect \centering{\includegraphics[height=6cm,
width=9cm]{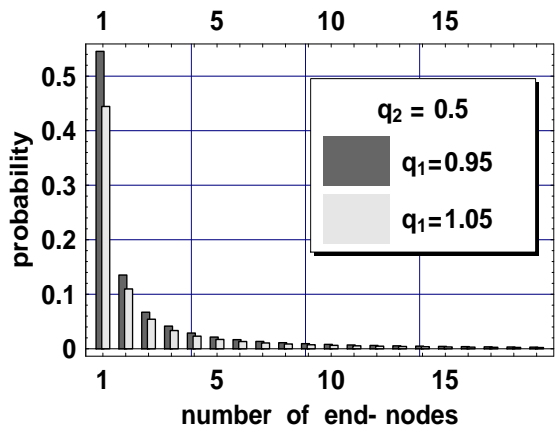}}\protect
\caption{\footnotesize{Dependence of probabilities $p_{n}^{(0)}$
on $n$ at $q_{2}=0.5$}} \label{fig4}
\end{figure}
\noindent i.e. the probability that the number of end-nodes in a
supercritical random tree is infinite and is given by
$2(q_{1}-1)/q_{2}$. At this point one has to remember that in Eq.
(\ref{12}) it was shown that $2(q_{1}-1)/q_{2}$, at the same time,
is equal to the probability of finding infinite number of nodes in
a supercritical random tree. At the first sight the result of this
comparison is surprising, but considering that the statement
expresses only the equality of measures characterizing sets of
nodes and end-nodes in an {\em infinite tree}, it is far not
unexpected. By expanding $g_{0}(z)$ in power series at $z=0$ one
obtains
\begin{equation}
\label{15} p_{n}^{(0)} = \frac{1}{\sqrt{\pi}}\;f_{0}
\frac{\Gamma(n-1/2)}{\Gamma(n+1)}\;
\frac{(4\;f_{0}\;f_{2})^{n-1}}{(1-f_{1})^{2n-1}}, \;\;\;\; n > 0.
\end{equation}

\begin{figure} [ht!]
\protect \centering{\includegraphics[height=6cm,
width=9cm]{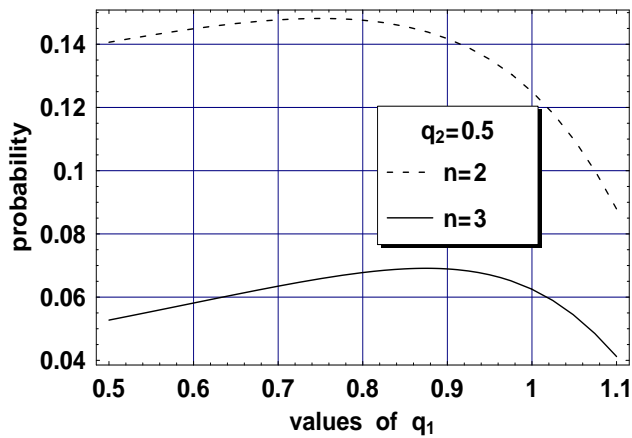}}\protect
\caption{\footnotesize{Dependence of probabilities $p_{n}^{(0)}$
on $q_{1}$ at $n=2, 3$ and $q_{2}=0.5$}} \label{fig5}
\end{figure}

\begin{figure} [ht]
\protect \centering{\includegraphics[height=8cm,
width=12cm]{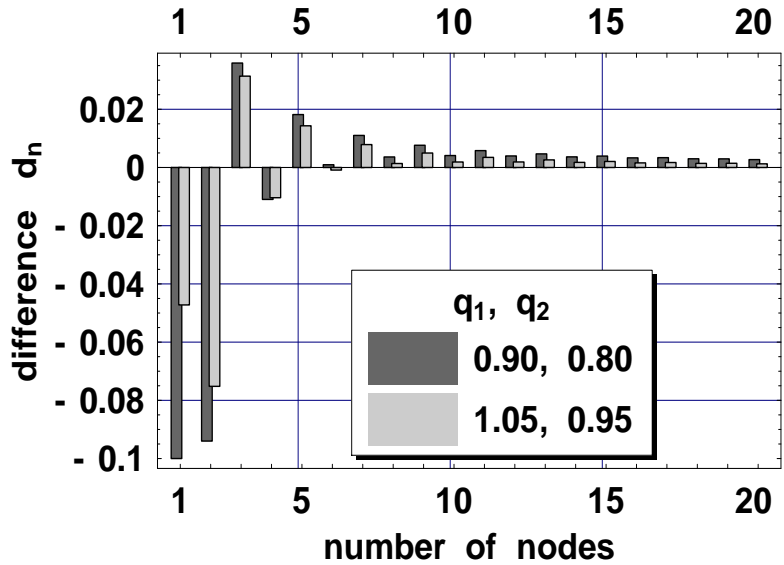}}\protect
\caption{\footnotesize{Variation of the difference $d_{n} = p_{n}
- p_{n}^{(0)}$ with $n$ in the case of sub- and supercritical
stationary random trees}} \label{fig6}
\end{figure}

In Fig. \ref{fig4} we can see how the probability of finding $n$
end-nodes in a stationary random tree depends on $n$ when
$q_{1}=0.95, \;\; q_{2}=0.5$ and $q_{1}=1.05, \;\; q_{2}=0.5$,
respectively. It is remarkable that the probabilities are
decreasing very rapidly with increasing $n$, i.e. there are very
few among the stationary trees containing a large number of
end-nodes. The dependence of $p_{2}^{(0)}$ and $p_{3}^{(0)}$ on
$q_{1}$ is shown in Fig. \ref{fig5} at $q_{2}=0.5$. One can
observe the formation of a maximum in both curves. The sites of
maxima are lying in the interval $0 << q_{1} <1$.

\begin{table}[ht!]
\begin{center}
Table $2$: {\footnotesize Probabilities $p_{n}$ and $p_{n}^{(0)}$
for small $n$}
\end{center}
\begin{center}
\begin{tabular}{|c|c|c|c|c|} \hline
$n$ & 1 & 2 & 3 & 4 \\ \hline $p_{n}$ & $f_{0}$ & $f_{0}f_{1}$ &
$f_{0} \left(f_{1}^{2} + f_{0}f_{2}\right)$ &  $f_{0}\;f_{1}
\left(f_{1}^{2} + 3f_{0}f_{2}\right)$ \\ \hline $p_{n}^{(0)}$ &
$f_{0}/(1-f_{1})$ & $f_{0}^{2}f_{2}/(1-f_{1})^{3}$ & $2f_{0}^{3}
f_{2}^{2}/(1-f_{1})^{5}$ & $5f_{0}^{4} f_{2}^{3}/(1-f_{1})^{7}$
\\ \hline
\end{tabular}
\label{tab2}
\end{center}
\end{table}

In order to compare the probabilities  $p_{n}$ and $p_{n}^{(0)}$
the difference $d_{n} = p_{n} - p_{n}^{(0)}$ has been calculated.
For small values of $n$ one can obtain $d_{n} < 0$. Fig.
\ref{fig6} shows the dependence of $d_{n}$ on $n$ in sub- and
supercritical stationary random trees. For the sake of simple
comparison of probabilities $p_{n}$ and $p_{n}^{(0)}$ for small
$n$ values Table $2$ has been compiled.

\subsection{Geometric distribution of $\nu$}

Let us investigate now the size distribution of stationary random
trees when $\nu$ is of geometric distribution, i.e
\begin{equation}
\label{16} q(z) = \frac{1}{1+q_{1}(1-z)}.
\end{equation}
It can be immediately seen  that the equation
\[  g(z) = \frac{z}{1 + q_{1}[1 - g(z)]} \]
has to be solved and
\[ g_{1,2}(z) =
\frac{1}{2q_{1}}\;\left[1+q_{1} \pm \sqrt{(1+q_{1})^{2} -
4q_{1}\;z}\right] \] are the two solutions. Since $g_{1}(0) = 1 +
1/q_{1} > 1$ the root
\begin{equation} \label{17} g_{2}(z) = g(z) =
\frac{1}{2q_{1}}\;\left[1+q_{1} - \sqrt{(1+q_{1})^{2} -
4q_{1}\;z}\right]
\end{equation}
has to be chosen as probability generating function for
$\tilde{\mu}$. Clearly,
\begin{equation}
\label{18} \lim_{z \uparrow 1} g(z) = \left\{ \begin{array}{ll}
1, & \mbox{if $q_{1} \leq 1$,} \\
\mbox{ } & \mbox{ } \\
1/q_{1}, & \mbox{if $q_{1} > 1$,} \end{array} \right.
\end{equation}
i.e. the probability to find infinite number of nodes in a
supercritical stationary random tree is $w_{\infty} = 1-1/q_{1}$.

By using the power series of $g(z)$ it can be shown that the
probability of finding $n$ nodes in a stationary tree is nothing
else, but
\begin{equation}
\label{19} p_{n} =
\frac{1}{\sqrt{\pi}}\;\frac{\Gamma(n-1/2)}{\Gamma(n+1)}\;
\frac{(4q_{1})^{n-1}}{(1+q_{1})^{2n-1}}.
\end{equation}

\begin{figure} [ht]
\protect \centering{\includegraphics[height=8cm,
width=12cm]{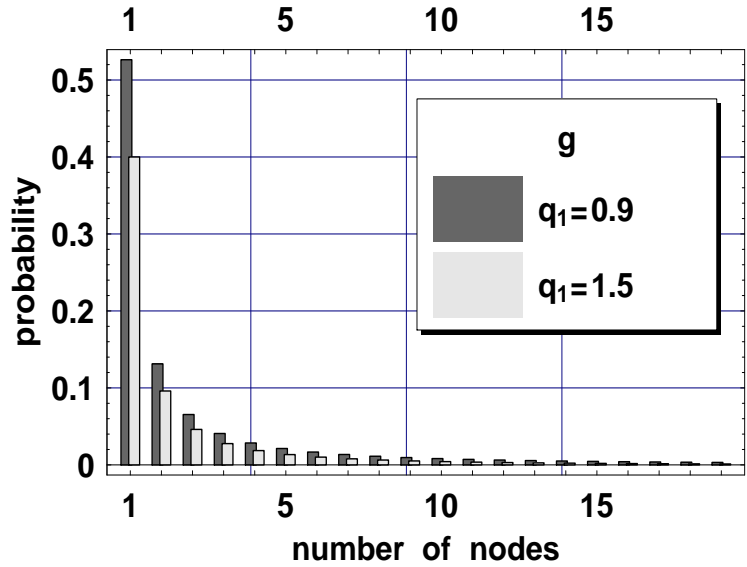}}\protect
\caption{\footnotesize{Probabilities to find $n=1, 2, \ldots$
nodes in a "very old" ($t=\infty$) tree when the distribution of
$\nu$ is geometric. The dark and light bars correspond to
subcritical and slightly supercritical tree evolutions,
respectively.}} \label{fig7}
\end{figure}

Fig. \ref{fig7} shows the dependence of the probability $p_{n}$ on
$n$. The dark bars denote probabilities to find $n=1, 2, \ldots$
nodes in a stationary tree produced in a subcritical ($q_{1}=0.9$)
evolution. At the same time, the light bars are related to
probabilities that $n=1, 2, \ldots$ nodes can be found in a
stationary tree which was produced by a strongly supercritical
($q_{1}=1.5$) evolution.

We underline again that ${\bf E}\{\tilde{\mu}\}$ exists and can be
calculated from Eq. (\ref{17}) only if $q_{1} < 1$.

The probability distribution of the number of end-nodes in
stationary random trees has some new qualitative features in
comparison with those obtained in the case $\nu \in {\bf t}$. By
using Eq. (\ref{7}) and taking into account the generating
function (\ref{16}) we obtain
\[ g_{0}(z) = \frac{1}{1 + q_{1}[1-g_{0}(z)]} -
\frac{1-z}{1+q_{1}},  \] and it is elementary to show that the
solution
\[ g_{0}(z) = \frac{1}{2 q_{1}(1+q_{1})}
\;\left[1 + q_{1} + q_{1}^{2} + q_{1}\;z - r(q_{1}, z) \right],\]
where
\[ r(q_{1}, z) = \sqrt{(1 + q_{1} + q_{1}^{2} + q_{1}\;z)^{2} - 4
q_{1}(1+q_{1})^{2}\;z},  \] is such that the requirement $g_{0}(1)
= 1$, if $q_{1} < 1$ is met. Expanding $g_{0}(z)$ into power
series around $z=0$ one finds~\footnote{The aim of the following
elementary consideration is to give help to understand how the
probability $p_{n}^{(0)}$ has been calculated. $r(q_{1}, z)$ can
be rewritten into the form:
\[ r(q_{1}, z) = q_{1}\;\sqrt{(z-z_{1})(z-z_{2})}, \]
where
\[ z_{1} = 1 + \frac{(1+q_{1})^{2}}{q_{1}} +
2\;\frac{1+q_{1}}{\sqrt{q_{1}}}, \] and
\[ z_{2} = 1 + \frac{(1+q_{1})^{2}}{q_{1}} -
2\;\frac{1+q_{1}}{\sqrt{q_{1}}}. \] Introducing the notations
$X(q_{1}) = 1/z_{1}$ and $Y(q_{1}) = 1/z_{2}$ one obtains \[
g_{0}(z)= \frac{1}{2
q_{1}(1+q_{1})}\left\{1+q_{1}+q_{1}^{2}+q_{1}z -
q_{1}\sqrt{z_{1}z_{2}}\;\sqrt{[1-X(q_{1})z][1-Y(q_{1})z]}\right\},\]
and this formula is used to get the power series of $g_{0}(z)$.}
that
\begin{equation}
\label{20} p_{n}^{(0)} = \left\{ \begin{array}{ll} \frac{1}{2
q_{1}(1+q_{1})}\left\{q_{1} + (1+q_{1}+q_{1}^{2})\;C_{1}[X(q_{1})
+ Y(q_{1})]\right\}, & \mbox{if $n=1$,} \\
\mbox{} & \mbox{} \\
\frac{1 + q_{1} + q_{1}^{2}}{2 q_{1}(1+q_{1})}
\;\left\{C_{n}\left[X(q_{1})^{n} + Y(q_{1})^{n}\right] -
Z_{n}(q_{1})\right\}, & \mbox{if $n > 1$,}
\end{array} \right.
\end{equation}
where
\[ C_{n} =
\frac{1}{2\sqrt{\pi}}\frac{\Gamma(n-1/2)}{\Gamma(n+1)}, \] while
\[ X(q_{1}) = \frac{q_{1}}{q_{1} + (1+q_{1})\;(1 +
\sqrt{q_{1}})^{2}}, \;\;\;\;\; Y(q_{1}) = \frac{q_{1}}{q_{1} +
(1+q_{1})\;(1 - \sqrt{q_{1}})^{2}} \] and
\[ Z_{n}(q_{1}) = \sum_{k=1}^{n-1} C_{k} C_{n-k}
X(q_{1})^{k}\;Y(q_{1})^{n-k}. \] It is relevant to note that
\[ \sum_{n=1}^{\infty} p_{n}^{(0)} = \left\{ \begin{array}{ll}
1, & \mbox{if $q_{1} \leq 1$,} \\
1/q_{1}, & \mbox{if $q_{1} > 1$.} \end{array} \right. \] In other
words, the probability to find infinite number of end-nodes in a
supercritical stationary tree is nothing else than $w_{\infty} = 1
- 1/q_{1}$.

\begin{figure} [ht!]
\protect \centering{\includegraphics[height=6cm,
width=9cm]{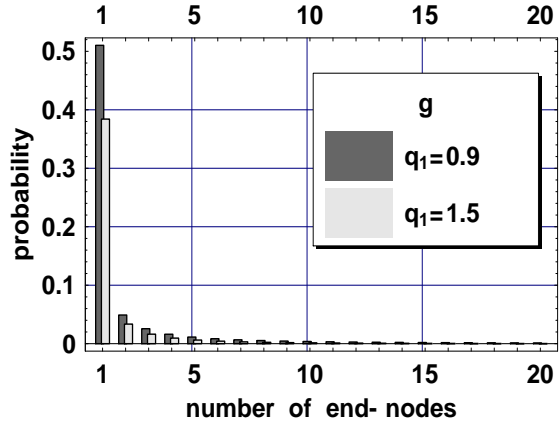}}\protect
\caption{\footnotesize{Probabilities to find $n=1, 2, \ldots$
end-nodes in a stationary tree when the distribution of $\nu$ is
geometric. The dark and light bars correspond to sub- and
supercritical evolutions, respectively.}} \label{fig8}
\end{figure}

Fig. \ref{fig8} shows the probabilities of finding $n=1, 2,
\ldots$ end-nodes in a stationary random tree when $\nu \in {\bf
g}$. We have seen in Fig. \ref{fig4} a similar bar chart but by
making a comparison between the two bar charts we can conclude
that the ratio of probability $p_{1}^{(0)}$ to $p_{2}^{(0)}$ is
much larger in the case of $\nu \in {\bf g}$ than in that of $\nu
\in {\bf t}$. In other words, if $\nu$ is of geometric
distribution then the probability of formation of rod like
stationary random trees is significantly greater than in the case
of $\nu \in {\bf t}$.

The structure of stationary random trees can be better visualized
by giving the formulas $p_{n}$ and $p_{n}^{(0)}$ for $n=1,2,3,4$
explicitly. Table $3$ contains these formulas.

\begin{table}[ht!]
\begin{center}
Table $3$: {\footnotesize Probabilities $p_{n}$ and $p_{n}^{(0)}$
for $n=1,2,3,4$}
\end{center}
\begin{center}
\begin{tabular}{|c|c|c|} \hline
$n$ & 1 & 2  \\ \hline $p_{n}$ & $1/(1+q_{1})$ &
$q_{1}/(1+q_{1})^{3}$ \\ \hline $p_{n}^{(0)}$ &
$(1+q_{1}/(1+q_{1}+q_{1}^{2})$ &
$q_{1}^{2}(1+q_{1})/(1+q_{1}+q_{1}^{2})^{3}$  \\ \hline
\end{tabular}
\end{center}
\begin{center}
\begin{tabular}{|c|c|c|} \hline
$n$ & 3 & 4  \\ \hline $p_{n}$ & $2q_{1}/(1+q_{1})^{5}$ &
$5q_{1}/(1+q_{1})^{7}$ \\ \hline $p_{n}^{(0)}$ &
$q_{1}^{3}(1+q_{1})h_{3}(q_{1})/(1+q_{1}+q_{1}^{2})^{5}$ &
$q_{1}^{4}(1+q_{1})h_{4}(q_{1})/(1+q_{1}+q_{1}^{2})^{7}$
\\ \hline
\end{tabular}
\label{tab3}
\end{center}
\end{table} \noindent
where
\[ h_{3}(q_{1}) = 1 + 3q_{1} + q_{1}^{2} \;\;\;\; \mbox{and}
\;\;\;\;  h_{4}(q_{1}) = 1 + 7q_{1} + 13q_{1}^{2} + 7q_{1}^{3} +
q_{1}^{4} \]

\begin{figure} [ht!]
\protect \centering{\includegraphics[height=6cm,
width=9cm]{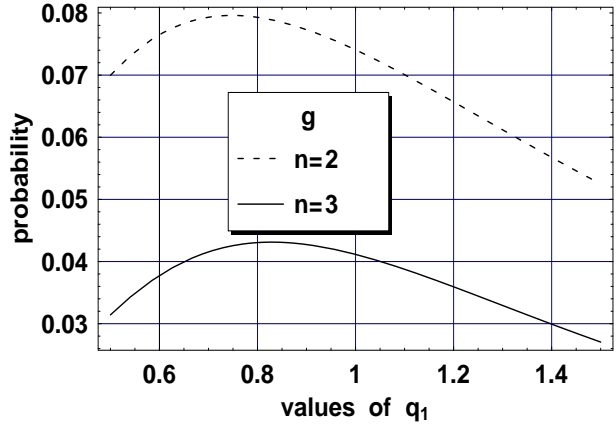}}\protect
\caption{\footnotesize{Dependence of probabilities $p_{n},
\;\;n=2,3$ on $q_{1}$ in the case of $\nu \in {\bf g}$.}}
\label{fig9}
\end{figure}

In the case of geometric distribution of $\nu$ the curves $p_{2}$
and $p_{3}$ versus $q_{1}$ are similar to those we have seen in
Fig. \ref{fig5} though the maxima are appearing at smaller values
of $q_{1}$ than in the case of $\nu \in {\bf t}$. The dependence
of probabilities $p_{n}, \; n = 2, 3$ on $q_{1}$ is presented in
Fig. \ref{fig9}.

\begin{figure} [ht!]
\protect \centering{\includegraphics[height=6cm,
width=9cm]{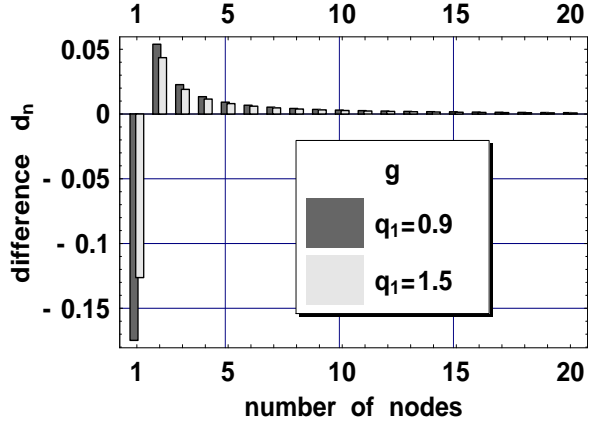}}\protect \caption{\footnotesize{Variation
of the difference $d_{n} = p_{n} - p_{n}^{(0)}$ with $n$ in the
case of sub- and supercritical stationary random trees when $\nu
\in {\bf g}$.}} \label{fig10}
\end{figure}

It seems to be instructive to show the variation of the
probability difference $d_{n} = p_{n} - p_{n}^{(0)}$ with $n$ in
the case of sub- and supercritical stationary random trees. In
Fig. \ref{fig10} one can see that $d_{n} > 0$ when $n > 1$.

\subsection{Poisson distribution}

In the sequel we would like to discuss  briefly the case when
$\nu$ has Poisson distribution with parameter $q_{1} > 0$. As
known, the generating function of $\nu$ is given by
\[ q(z) = {\bf E}\{z^{\nu}\} = e^{-(1-z)q_{1}} \]
and ${\bf E}\{\nu\} = {\bf D}^{2}\{\nu\} = q_{1}.$

It can be easily shown that the generating function  $g(z) = {\bf
E}\{z^{\tilde{\mu}}\}$ of the random variable $\lim_{t \rightarrow
\infty} \mu(t)  \stackrel{\mathrm{d}}{=} \tilde{\mu}$ should
satisfy the following equation:
\begin{equation}
\label{21} g(z) = z\;q[g(z)] = z\;e^{-q_{1}}\;e^{q_{1}g(z)}.
\end{equation}

Applying Theorem \ref{th1} to  Eq. (\ref{21}) one can formulate
the following statement:

\newpage

{\em If $z \uparrow 1$ and $0 < q_{1} \leq 1$, then the equation
$\tilde{g} = e^{-q_{1}}\;e^{q_{1}\tilde{g}}$, where $\tilde{g} =
\lim_{z \uparrow 1}g(z) = \sum_{n=0}^{\infty}p_{n}$, has only one
root in the interval $[0, 1]$, and that is $\tilde{g} = g_{1} =
1$, while if $q_{1} > 1$, then besides $g_{1}$ there is another
root in $[0, 1]$, namely $\tilde{g} = g_{2} < 1$.}

The consequence of this statement means that if  $q_{1} > 1$, then
the probability to find infinite number of nodes in a stationary
random tree in the case of $\nu \in {\bf p}$ is nothing else than
\[ w_{\infty} = 1 - g_{2} = 1 - \sum_{n=0}^{\infty} p_{n}. \]

\begin{figure} [ht!]
\protect \centering{\includegraphics[height=6cm,
width=9cm]{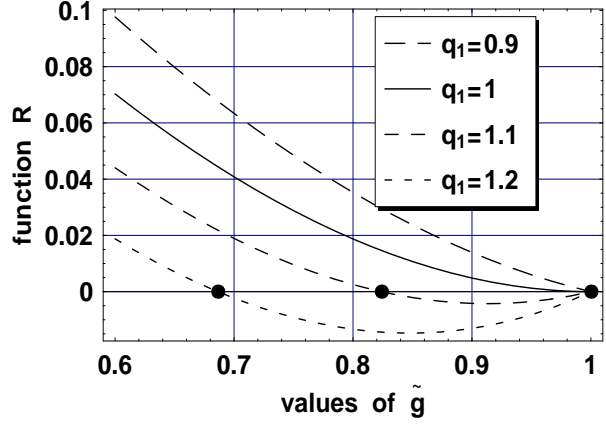}}\protect
\caption{\footnotesize{Illustration of the appearance of roots
$g_{2} < 1$ when $q_{1} = 1.1$ and $1.2$.}} \label{fig11}
\end{figure}

In Fig. \ref{fig11} we would like to illustrate the appearance of
roots smaller than $1$ of the equation $R = e^{-q_{1}}
\;e^{q_{1}\tilde{g}} - \tilde{g} = 0$. We can see, that the
equation $R = 0$ has only one trivial root in $[0,1]$, namely the
$g_{1}=1$, when $0 < q_{1} \leq 1$. The black points are referring
to roots due to four different values of $q_{1}$.

\begin{figure} [ht!]
\protect \centering{\includegraphics[height=6cm,
width=9cm]{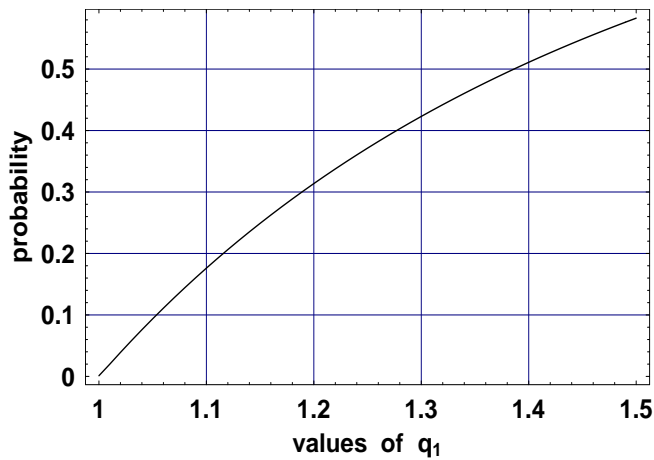}}\protect
\caption{\footnotesize{Dependence of probability $w_{\infty}$ on
$q_{1} > 1$.}} \label{fig12}
\end{figure}

The probability to find infinite number of nodes in a stationary
random tree has been calculated, and the $w_{\infty}$ versus
$q_{1}$ curve is plotted in Fig. \ref{fig12}. It is interesting to
note that $w_{\infty} > 0.5$ when $q_{1}=1.4$.

Now, we want to calculate the probabilities $p_{n}$ of finding
$n=1,2,\ldots$ nodes in a stationary random tree. Expanding the
expression \[ z\;\exp(-q_{1})\;\exp\left[-q_{1}
\sum_{n=1}^{\infty} p_{n}z^{n}\right] \] into  power series of $z$
at $z=0$ we can step-by-step compute the probabilities $p_{n},
\;\; n=1,2, \ldots$, in accordance with Eq. (\ref{21}). We obtain
that
\begin{equation}
\label{22} p_{n} =
C_{n}\;q_{1}\;\left[e^{-q_{1}}\;q_{1}\right]^{n},
\end{equation}
where $C_{n}, \;\; n=1, 2, \ldots$ are positive rational numbers.
The first seven of them are given in Table $4$.  It seems to be
hardly possible to obtain explicit an formula for $C_{n}$, but it
is an easy task to compose an algorithm for its computation.

\begin{table}[ht!]
\begin{center}
Table $4$: {\footnotesize Coefficients $C_{n}$ in  $p_{n}$ for
$n=1,2,3,4, 5$}
\end{center}
\begin{center}
\begin{tabular}{|c|c|c|c|c|c|c|c|} \hline
$n$ & 1 & 2 & 3 & 4 & 5 & 6 & 7  \\ \hline $C_{n}$ & $1$ & $1$ &
$3/2$ & $8/3$ & $125/24$ & $54/5$ & $16807/720$ \\ \hline
\end{tabular}
\label{tab4}
\end{center}
\end{table}

\begin{figure} [ht!]
\protect \centering{\includegraphics[height=6cm,
width=9cm]{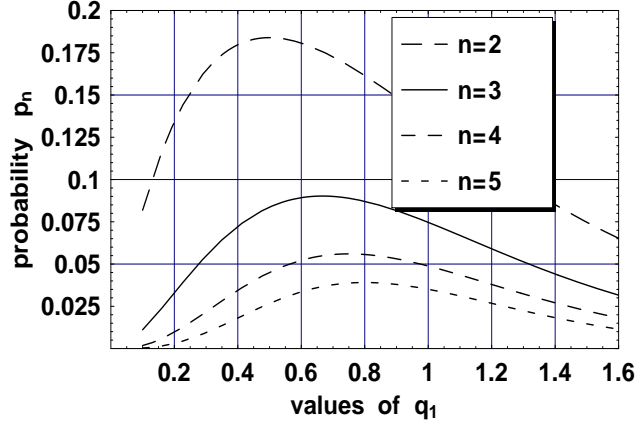}}\protect
\caption{\footnotesize{Dependence of probability $p_{n}$ on
$q_{1}$  for $n=2,3,4,5$.}} \label{fig13}
\end{figure}

It follows from Eq. (\ref{22} immediately that the probability
$p_{n}$ vs. $q_{1}$ has a maximum at \[ q_{1} = 1 - \frac{1}{n}.
\] The appearance of maxima is well seen in Fig. \ref{fig13} in
the cases of $n=2,3,4,5$.

\vspace{0.4cm}

Finally, we would like to discuss the {\em properties of
end-nodes} in stationary random trees evolved according to Poisson
distribution of $\nu$. In order to obtain the probabilities
$p_{n}^{(0)}\;\; n=1,2, \ldots$ we have to solve the equation
\[ \exp(-q_{1})\;\exp\left[q_{1}\;g_{0}(z)\right] = g_{0}(z) +
(1-z)\exp(-q_{1}), \] where
\[ g_{0}(z) = \sum_{n=1}^{\infty} p_{0}^{(0)} z^{n}. \]
By using a special expansion procedure we can get the
probabilities $\;p_{1}^{(0)},\;$ \; $\;p_{2}^{(0)}, \; \ldots$
step-by-step. Introducing the notations
\[ u(q_{1}) = (e^{q_{1}} - q_{1})^{-1},  \]
\[ \ell_{3}(q_{1}) = e^{q_{1}} + 2q_{1}, \]
\[ \ell_{4}(q_{1}) = e^{2q_{1}} + 8 e^{q_{1}} q_{1} + 6 q_{1}^{2}, \]
\[ \ell_{5}(q_{1}) = e^{3q_{1}} + 22 e^{2q_{1}} q_{1} + 58
e^{q_{1}} q_{1}^{2} + 24 q_{1}^{3}, \]
\[ \ell_{6}(q_{1}) = e^{4q_{1}} + 52 e^{3q_{1}} q_{1} + 328
e^{2q_{1}} q_{1}^{2} + 444 e^{q_{1}} q_{1}^{3} + 120 q_{1}^{4} \]
the first six probabilities are given by the following Eqs.:
\[ p_{1}^{(0)} = u(q_{1}), \;\;\;\;
p_{2}^{(0)} = \frac{1}{2} q_{1}^{2} [u(q_{1})]^{3}, \;\;\;\;
p_{3}^{(0)} = \frac{1}{6} q_{1}^{3} \ell_{3}(q_{1})
[u(q_{1})]^{5}, \]
\[ p_{4}^{(0)} = \frac{1}{24} q_{1}^{4} \ell_{4}(q_{1}) [u(q_{1})]^{7}, \;\;\;\;
p_{5}^{(0)} = \frac{1}{120} q_{1}^{5} \ell_{5}(q_{1})
[u(q_{1})]^{9},\] \[ p_{6}^{(0)} = \frac{1}{720} q_{1}^{6}
\ell_{6}(q_{1}) [u(q_{1})]^{11}. \]

\newpage

\begin{figure} [ht!]
\protect \centering{\includegraphics[height=6cm,
width=9cm]{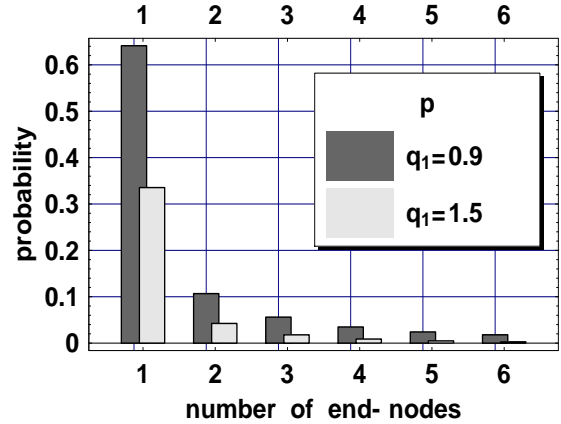}}\protect
\caption{\footnotesize{Probabilities to find $n=1,\ldots 6$
end-nodes in a stationary random tree when $\nu \in {\bf p}$.}}
\label{fig14}
\end{figure}

\begin{figure} [ht!]
\protect \centering{\includegraphics[height=6cm,
width=9cm]{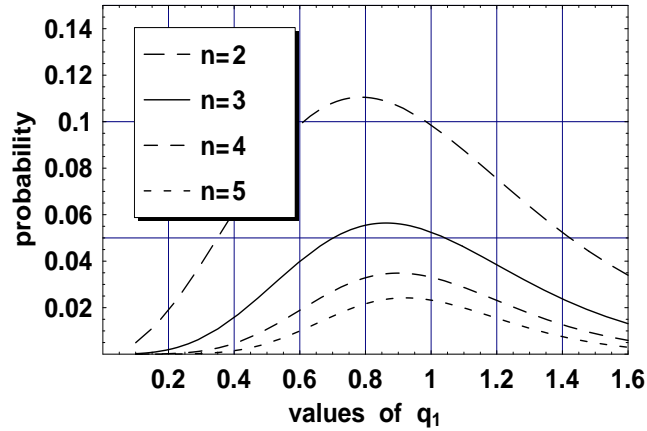}}\protect
\caption{\footnotesize{Dependence of probabilities $p_{n}^{(0)},
\;\; n=2,3,4,5$  on $q_{1}$ when $\nu \in {\bf p}$.}}
\label{fig15}
\end{figure}

By using these formulas we are able to show the dependence of
$p_{n}^{(0)}$ on $n$ and $q_{1}$, respectively. We see in Fig.
\ref{fig14} that the evolution process prefers the one end-node
structures, i.e. stationary random trees in which the probability
of finding $n > 1$ end-nodes is very small. The curves
$p_{n}^{(0)}, \;\; n=2,3,4,5$ versus $q_{1}$ plotted in Fig.
\ref{fig15} are clearly demonstrating that the sites of maxima of
probabilities are shifted to higher values of $q_{1}$ as the
number of nodes $n$ is increasing.

\section{Conclusions}

We have investigated the main properties of evolution of special
trees when the continuous time parameter tends to infinity. A tree
of infinite age is called {\em stationary} or simply "very old".
We have stated that if the limit relations
\[ \lim_{t \rightarrow \infty} {\mathcal P}\{\mu(t)=n|{\mathcal
S}_{0}\} = p_{n} \leq 1, \;\;\;\;\;\; \forall \; n = 1, 2, \ldots
\] are true, then  the random function $\mu(t)$ giving the
number of nodes in a tree at $t \geq 0$ {\em converges in
distribution} to a {\em random variable $\tilde{\mu}$} which
counts the number of nodes in a stationary tree.

For the generating function ${\bf E}\{z^{\tilde{\mu}}\} = g(z)$ a
simple equation has been derived, namely \[ g(z) = zq[g(z)],
\;\;\;\;\;\;\;\;\;\;\;\;\;\;\;\;\;\;\;\; \mbox{{\bf (a)}} \] where
$q(z)$ is nothing else than the generating function of the number
of new nodes $\nu$ produced by one dying node. It has been proved
that if $\nu$ has finite first and second factorial moments, i.e.
if ${\bf E}\{\nu\}=q_{1}$ and ${\bf E}\{\nu (\nu-1)\}=q_{2}$ are
finite, then $\sum_{n=1}^{\infty} p_{n} = 1$, if $q_{1} \leq 1$,
and $\sum_{n=1}^{\infty} p_{n} \leq 1$, if $q_{1} > 1$. It means,
if $q_{1} \leq 1$, then the probability $w_{\infty}$ of finding
infinite number of nodes in a stationary random tree is zero,
while if  $q_{1} > 1$, then that is larger than zero, namely,
$w_{\infty} = 1 - \sum_{n=1}^{\infty} p_{n} \leq 1$.

Here, one has to note that the expectation value ${\bf
E}\{\tilde{\mu}\}$ exists only if the tree evolution is
subcritical, i.e. if $q_{1} < 1$. In the case of $q_{1} > 1$, i.e.
in supercritical evolution it has been shown that \[\lim_{t
\rightarrow \infty} {\bf E}\{\mu(t)\;\exp[-(q_{1}-1) Qt]\} =
q_{1}/(q_{1}-1),\] while if $q_{1}=1$, then $\lim_{t \rightarrow
\infty} {\bf E}\{\mu(t)/Qt\} = 1$.

It has been shown also that the generating function $g_{0}(z)$ of
the number of end-nodes $\tilde{\mu}_{0}$ in a stationary random
tree has to satisfy the equation \[ g_{0}(z) = q[g_{0}(z)] -
(1-z)f_{0}, \;\;\;\;\;\;\;\;\;\;\;\;\;\;\;\;\;\;\;\; \mbox{{\bf
(b)}}\]  which can be used to study end-node properties. When
$q(z)={\bf E}\{z^{\nu}\}$ is known then by using appropriate
method for comparison of coefficients of $z^{n}$ in both sides of
equations {\bf (a)} and  {\bf (b)} we could  calculate
step-by-step the probabilities $p_{1}, p_{1}^{(0)},\; p_{2},
p_{2}^{(0)}, \ldots$ to find $n=1, 2, \ldots$ nodes and end-nodes,
respectively, in a stationary random tree.

For the calculations, we have chosen three different distributions
of $\nu$. The generating functions of these distributions  are
given by the following formulas:
\[ q(z) = \left\{ \begin{array}{ll}
1 + q_{1}(z-1) + \frac{1}{2} q_{2}(z-1)^2, & \mbox{if $\nu \in
{\bf t}$,} \\
\mbox{} & \mbox{ } \\
1/[1 + q_{1}(z-1)], & \mbox{if $\nu \in {\bf g}$,} \\
\mbox{} & \mbox{ } \\
e^{q_{1}(z-1)}, & \mbox{if $\nu \in {\bf p}$.}
\end{array} \right. \]

Analyzing the results of numerical calculations the first
impression is that the qualitative properties of stationary random
trees depend hardly on the character of distribution of $\nu$. We
have seen that in all cases the probability to find $n=1$ node (or
end-node) in a stationary tree is significantly larger then to
find $n>1$ nodes. One can conclude that in the evolution process
the formation of a rod-like stationary random tree is much more
probable than that with many branches.

We have found special behavior in the dependence of $p_{n}$ on $n$
only in the case of $\nu \in {\bf t}$ when ${\mathcal
P}\{\nu=1\}=f_{1}$ is smaller than a critical value. The
appearance of the oscillation of $p_{n}$ versus $n$ is consequence
of the following trivial statement: when $f_{1}=0$ then $p_{2j} =
0, \;\; j=1,2,\ldots\;$.

It has been demonstrated  that the probabilities $p_{n}$ and
$p_{n}^{(0)}$ versus $q_{1}$ show a maximum the location of which
is increasing with $n$ but remains always smaller than $1$. This
property is best seen in the case of Poisson distribution of
$\nu$.

\end{document}